\documentclass[traditabstract]{aa}

\usepackage{epsfig,graphicx,natbib}

\def\Tcmb{\hbox{$T_\mathrm{CMB}$}}

\def\Tkin{\hbox{$T_\mathrm{kin}$}}

\def\nH2{\hbox{$n_\mathrm{H_2}$}}
\def\kms{\hbox{km\,s$^{-1}$}}
\def\PKS1830{\hbox{PKS\,1830$-$211}}
\def\cm-2{\hbox{cm$^{-2}$}}
\def\fH2{\hbox{$f_{\rm H2}$}}
\def\xe{\hbox{$x_e$}} 
\def\nH{\hbox{$n_{\rm H}$}}
\def\zzH{\hbox{$\zeta_{\rm H}$}}

\DeclareSymbolFont{matha}{OML}{txmi}{m}{it}  
\DeclareMathSymbol{\varv}{\mathord}{matha}{118}

\begin{document}

\title{OH$^+$ and H$_2$O$^+$ absorption toward \PKS1830}

\author{S.~Muller \inst{1}
\and H.\,S.\,P.~M{\"u}ller \inst{2}
\and J.\,H.~Black \inst{1}
\and A.~Beelen \inst{3}
\and F.~Combes \inst{4}
\and S. Curran \inst{5}
\and M.~G\'erin \inst{6}
\and M.~Gu\'elin \inst{7,6}
\and C.~Henkel \inst{8,9}
\and S. Mart\'in \inst{10}
\and S.~Aalto \inst{1}
\and E.~Falgarone \inst{6}
\and K.\,M.~Menten \inst{8}
\and P. Schilke \inst{2}
\and T. Wiklind \inst{11}
\and M.\,A.~Zwaan \inst{12}
}
\institute{Department of Earth and Space Sciences, Chalmers University of Technology, Onsala Space Observatory, SE-43992 Onsala, Sweden
\and I.~Physikalisches Institut, Universit{\"a}t zu K{\"o}ln, Z{\"u}lpicher Str. 77, 50937 K{\"o}ln, Germany
\and Institut d'Astrophysique Spatiale, B\^at. 121, Universit\'e Paris-Sud, 91405 Orsay Cedex, France
\and Observatoire de Paris, LERMA, CNRS, 61 Av. de l'Observatoire, 75014 Paris, France
\and School of Chemical and Physical Sciences, Victoria University of Wellington, PO Box 600, Wellington 6140, New Zealand 
\and LERMA, Observatoire de Paris \& Ecole Normale Sup\'erieure, Paris, France
\and Institut de Radioastronomie Millim\'etrique, 300, rue de la piscine, 38406 St Martin d'H\`eres, France 
\and Max-Planck-Institut f\"ur Radioastonomie, Auf dem H\"ugel 69, D-53121 Bonn, Germany
\and Astron. Dept., King Abdulaziz University, P.O. Box 80203, Jeddah, Saudi Arabia
\and European Southern Observatory, Alonso de C\'ordova 3107, Vitacura Casilla 763 0355, Santiago, Chile
\and Department of Physics, Catholic University of America, 620 Michigan Ave NE, Washington DC 20064
\and European Southern Observatory, Karl-Schwarzschild-Str. 2, 85748 Garching b. M\"unchen, Germany
}

\date {Received / Accepted}

\titlerunning{OH$^+$ and H$_2$O$^+$ toward \PKS1830}
\authorrunning{}

\abstract{We report the detection of OH$^+$ and H$_2$O$^+$ in the $z=0.89$ absorber toward the lensed quasar \PKS1830. The abundance ratio of OH$^+$ and H$_2$O$^+$ is used to quantify the molecular hydrogen fraction (\fH2) and the cosmic-ray ionization rate of atomic hydrogen (\zzH) along two lines of sight, located at $\sim 2$~kpc and $\sim 4$~kpc to either side of the absorber's center. 
The molecular fraction decreases outwards, from $\sim 0.04$ to $\sim 0.02$, comparable to values measured in the Milky Way at similar galactocentric radii. For \zzH, we find values of $\sim 2 \times 10^{-14}$~s$^{-1}$ and $\sim 3 \times 10 ^{-15}$~s$^{-1}$, respectively, which are slightly higher than in the Milky Way at comparable galactocentric radii, possibly due to a higher average star formation activity in the $z=0.89$ absorber. The ALMA observations of OH$^+$, H$_2$O$^+$, and other hydrides toward \PKS1830 reveal the multi-phase composition of the absorbing gas. Taking the column density ratios along the southwest and northeast lines of sight as a proxy of molecular fraction, we classify the species ArH$^+$, OH$^+$, H$_2$Cl$^+$, H$_2$O$^+$, CH, and HF as tracing gases increasingly more molecular. Incidentally, our data allow us to improve the accuracy of H$_2$O$^+$ rest frequencies and thus refine the spectroscopic parameters.}
\maketitle

\section{Introduction}

Hydrides, that is, molecules or molecular ions containing a single heavy element with one or more hydrogen atoms, are formed in the interstellar medium (ISM) from the initial chemical reactions starting from atomic gas, making them fundamental to interstellar chemistry. Their rotational spectrum starting at submm/FIR wavelengths (a few exceptions, such as H$_2$S and H$_2$Cl$^+$, have ground-state transitions in the millimeter band)
makes them difficult or even impossible to study from the ground. For this reason, many hydrides have only recently been detected in the ISM (see the review by \citealt{ger16}).

The two species of interest here, hydroxylium (OH$^+$) and oxidaniumyl (H$_2$O$^+$), were first detected in the Galactic interstellar medium by \cite{wyr10} with the Atacama Pathfinder EXperiment (APEX) telescope, and by \cite{oss10} with the Herschel Space Observatory, respectively. These two species, and the third oxygen-bearing ion in the family, hydronium (H$_3$O$^+$), are formed from O$^+$ by successive hydrogen abstraction reactions with H$_2$. There can also be a significant contribution from the H$_3^+$+O reaction, in any place where the molecular fraction is relatively high. OH$^+$ and H$_2$O$^+$ are destroyed by dissociative recombination with electrons and by reactions with H$_2$. H$_3$O$^+$ does not react with H$_2$ and is mostly destroyed by dissociative recombination. The relative abundances of these species are thus controlled by the molecular hydrogen fraction and cosmic-ray ionization rate \citep{hol12}, providing a powerful diagnostic of these two quantities. \cite{ind15} analyzed Herschel observations of OH$^+$, H$_2$O$^+$, and H$_3$O$^+$ in multiple lines of sight toward bright submillimeter continuum sources in the Galaxy. Their results confirm that OH$^+$ and H$_2$O$^+$ are primarily tracing gases with a low molecular hydogen fraction ($\fH2 = 2 \times N({\rm H}_2) / [N({\rm H})+2 \times N({\rm H}_2)]$) of a few percent. They also determined the cosmic-ray ionization rate of atomic hydrogen (\zzH) across the Galactic disk and found an average $\zzH \sim 2 \times 10^{-16}$~s$^{-1}$ that shows little variation over the disk outside of Galactocentric radii of 5~kpc. Closer to the Galactic center, \zzH\ is found to increase by up to two orders of magnitude in the central region itself. To date, there are only a few observations of OH$^+$ and/or H$_2$O$^+$ in extragalactic sources (e.g., \citealt{vdwer10, yan13, gon13}), including detections at very high redshift \citep{wei13, rie13}. \cite{gon13} report ionization rates toward Arp\,220 and NGC\,4418 comparable to, or possibly higher than, values in the Galactic center.

Although the Herschel Space Observatory is no longer in operation, it is now possible to extend the studies of these hydrides to other galaxies, e.g., with the Atacama Large Millimeter/submillimeter Array (ALMA). The molecular-line rich absorber at $z=0.89$ located in front of the quasar \PKS1830\ is certainly a prime target in this prospect. \PKS1830\ is lensed by the absorber, a nearly face-on typical spiral galaxy (e.g., \citealt{wik98,win02,koo05}). Molecular absorption is seen along the two independent lines of sight toward the southwest and northeast images of the quasar, at galactocentric radii of $\sim 2$~kpc and $\sim 4$~kpc in opposite directions from the center of the absorbing galaxy, respectively. The apparent size of the lensed images of the quasar is a fraction of a milliarcsecond at mm wavelengths (\citealt{jin03}). The volume of gas seen in absorption in each line of sight is thus roughly enclosed in a cylinder with a base of $< 1$~pc in diameter and a depth of a few tens to hundreds of pc, depending on the inclination ($i=17^\circ-32^\circ$, \citealt{koo05}) and the thickness of the absorber's disk. More than forty molecular species have been detected toward the southwest image, where the H$_2$ column density is $\sim 2 \times 10^{22}$~\cm-2\ (e.g., \citealt{mul11,mul14a}). In contrast, only about a dozen species have been detected toward the northeast image, where the molecular gas has a lower column and is more diffuse. Several submm lines of hydrides have already been observed toward \PKS1830\ with ALMA, such as those for CH, H$_2$O, NH$_2$, NH$_3$ \citep{mul14a}, H$_2$Cl$^+$ \citep{mul14b}, ArH$^+$ \citep{mul15}, as well as HF \citep{kaw16}.

\section{Observations} \label{sec:obs}

We observed the OH$^+$ $N$=1--0 $J$=1--0 line (two hyperfine components at a rest frequency near 909~GHz) redshifted into ALMA band~8 and the {\em p}-H$_2$O$^+$ $J_{\rm K_a,K_c}$=$1_{1,0}$--$1_{0,1}$ transition (four fine-structure components at rest frequencies 604, 607, 631, and 634~GHz) redshifted into band~7. We complement this dataset with contemporary observations of some other hydrides, collected at several epochs in Cycles 1, 2, and 3, between 2014 and 2016 (see Table~\ref{tab:obs}). The array was always in a configuration providing a synthesized beam resolution better than 0.5$''$, sufficient to resolve the two point-like lensed images of \PKS1830, separated by 1$''$. 

The correlator was set up with spectral basebands of 1.875~GHz, providing a velocity resolution of $\sim 0.6$--1.0~\kms. For each epoch, the bandpass response of the antennas was calibrated from observations of the bright quasar J1924$-$292, with J1832$-$2039 being used for primary gain calibration. The data calibration was done with the CASA\footnote{http://casa.nrao.edu/} package. We further improved the data quality using self-calibration on the bright continuum of \PKS1830. The spectra were then extracted toward both lensed images of \PKS1830, using the CASA-python task UVMULTIFIT \citep{mar14} in order to fit a model of two point sources to the visibilities.

\begin{figure}[h] \begin{center}
\includegraphics[width=8.8cm]{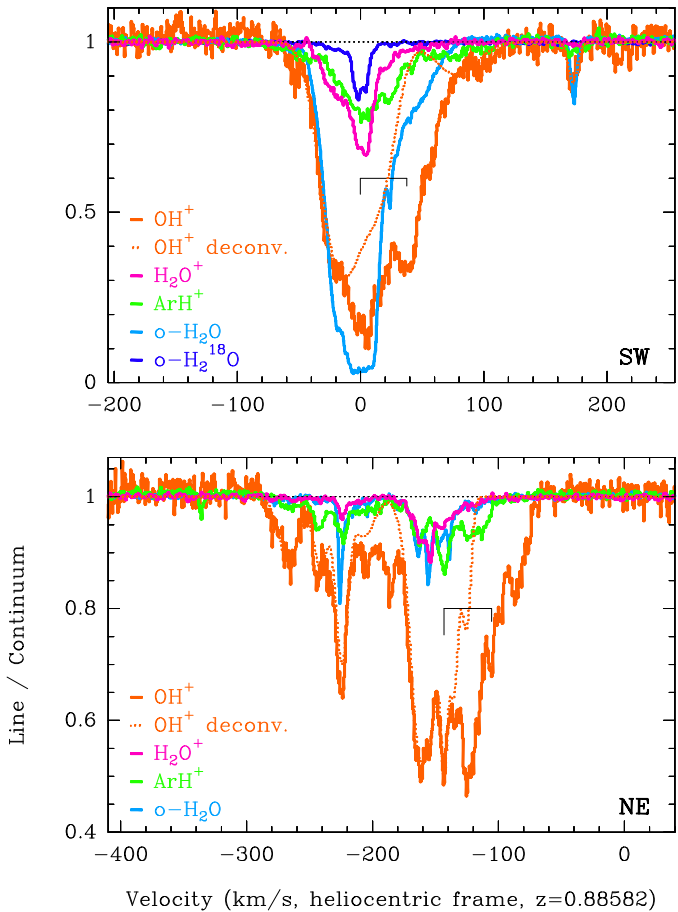}
\caption{Spectra of OH$^+$ and H$_2$O$^+$ (only the 607~GHz transition is shown) toward \PKS1830\ southwest ({\em top}) and northeast ({\em bottom}) images. The hyperfine structure (hfs) for OH$^+$ is indicated, with the relative strengths expected when the sublevels are populated in proportion to their statistical weights. The spectra corresponding to the strongest OH$^+$ hfs component, after hfs deconvolution, are shown in dotted lines. Spectra of other species are shown for comparison.}
\label{fig:spec-OH+-H2O+}
\end{center} \end{figure}

The absorption spectra of OH$^+$ (909~GHz) and H$_2$O$^+$ (the strongest fine-structure transition at 607~GHz) toward \PKS1830\ are shown in Fig.~\ref{fig:spec-OH+-H2O+}, together with spectra of some other species for comparison.

\begin{table}[ht]
\caption{ALMA observations discussed in this paper.} \label{tab:obs}
\begin{center} \begin{tabular}{ccl}
\hline \hline
Species & Approximate rest frequency & \multicolumn{1}{c}{Date} \\
        & (GHz)      &  \\
\hline
OH$^+$  & 909        & 2015 June 06 \\
H$_2$O$^+$ & 604, 607, 631 & 2014 May 03 \\
           & --              &  2014 May 06 \\
           & --              &  2014 June 30 \\
H$_2$O$^+$ / ArH$^+$           & 634 / 617           &  2015 May 19 \\
H$_2$O / CH & 557 / 532 / 536  & 2014 May 05 \\
       & --  & 2014 July 19 \\
       & --  & 2016 March 05 \\
H$_2^{18}$O & 547 & 2014 May 05 \\
           & --    & 2014 July 18 \\
\hline
\end{tabular}
\end{center} \end{table}

\section{Spectroscopy}

\subsection{OH$^+$}

The OH$^+$ radical has a $^3\Sigma ^-$ ground vibronic state with two unpaired electrons. The spin-spin coupling causes fairly large fine structure splittings. The spin of the H nucleus causes additional hyperfine structure (HFS) splitting. The transition frequencies of the $N$=1--0 transition were recorded by \cite{bek85} with $\sim 1$~MHz accuracy. The THz spectrum of OH$^+$ has recently been revised by \cite{mar16} and frequencies of the $N$=1--0 rotational transitions are predicted with an accuracy better than 1~MHz (i.e., $<0.3$~\kms). The $J$=0--1 fine structure (FS) component near 909~GHz is the weakest FS component and thus least prone to be affected by optical depth effects. Its HFS splitting (two components) is slightly larger than 100~MHz. The dipole moment, $\mu_{\rm OH^+}=2.3$~D, is from a quantum chemical calculation \citep{wer83}. The spectroscopic details were obtained from the Cologne Database for Molecular Spectroscopy (CDMS) \footnote{http://www.astro.uni-koeln.de/cdms/}.

\subsection{H$_2$O$^+$}

The H$_2$O$^+$ radical is an asymmetric top molecule with one unpaired electron which results in a $^2B_2$ ground vibronic state. The two H nuclei give rise to HFS for rotational levels with $K_a$+$K_c$ being even ($I_{\rm{tot}}=1$; \textit{ortho}) whereas those with $K_a$+$K_c$ being odd ($I_{\rm{tot}}=0$; \textit{para}) do not display HFS. The dipole moment is $\mu_{\rm H_2O^+}=2.4$~D \citep{wei89}. The initial spectroscopic data were taken from the CDMS, where it was noted that existing laboratory measurements are not of microwave precision. As shown in Fig.~\ref{fig:spec-H2O+-CDMS-revised}(left) and Table~\ref{H2O+_freqs}, the frequencies provided in the CDMS agree with our observations well enough for identification, but are not adequate for detailed kinematical analysis of the line profiles. Therefore, we decided to revise the rest frequencies of the four FS components of {\em p}-H$_2$O$^+$ observed with ALMA (Sec.~\ref{sec:evaluateH2O+freq}) and update the H$_2$O$^+$ entry in the CDMS (see Appendix~\ref{sec:LabSpectroH2O+}).

\section{Results and discussion}

\subsection{Observed spectra}

The OH$^+$ absorption profiles are deep in both lines of sight toward \PKS1830. They do not seem to be saturated, as there is no sign of a flattening of the profile bottom, as is the case  for H$_2$O, for example. Toward the southwest image, a broad ($\sim 100$~\kms, full width at zero power, FWZP) and deep (optical depth, $\tau \geq 3$) absorption component is seen near $v=0$~\kms, and the weak and narrow velocity component at $v=+170$~\kms\ (see \citealt{mul11,mul14a}) is also detected. Toward the northeast image, absorption occurs in a series of narrow components throughout the velocity interval between $-$300 and $-$100~\kms. The maximum absorption depth toward the northeast image implies that the continuum covering factor is larger than 50\%. This is the deepest molecular absorption detected along this line of sight, so far.

\begin{figure*}[ht!] \begin{center}
\includegraphics[width=8.8cm]{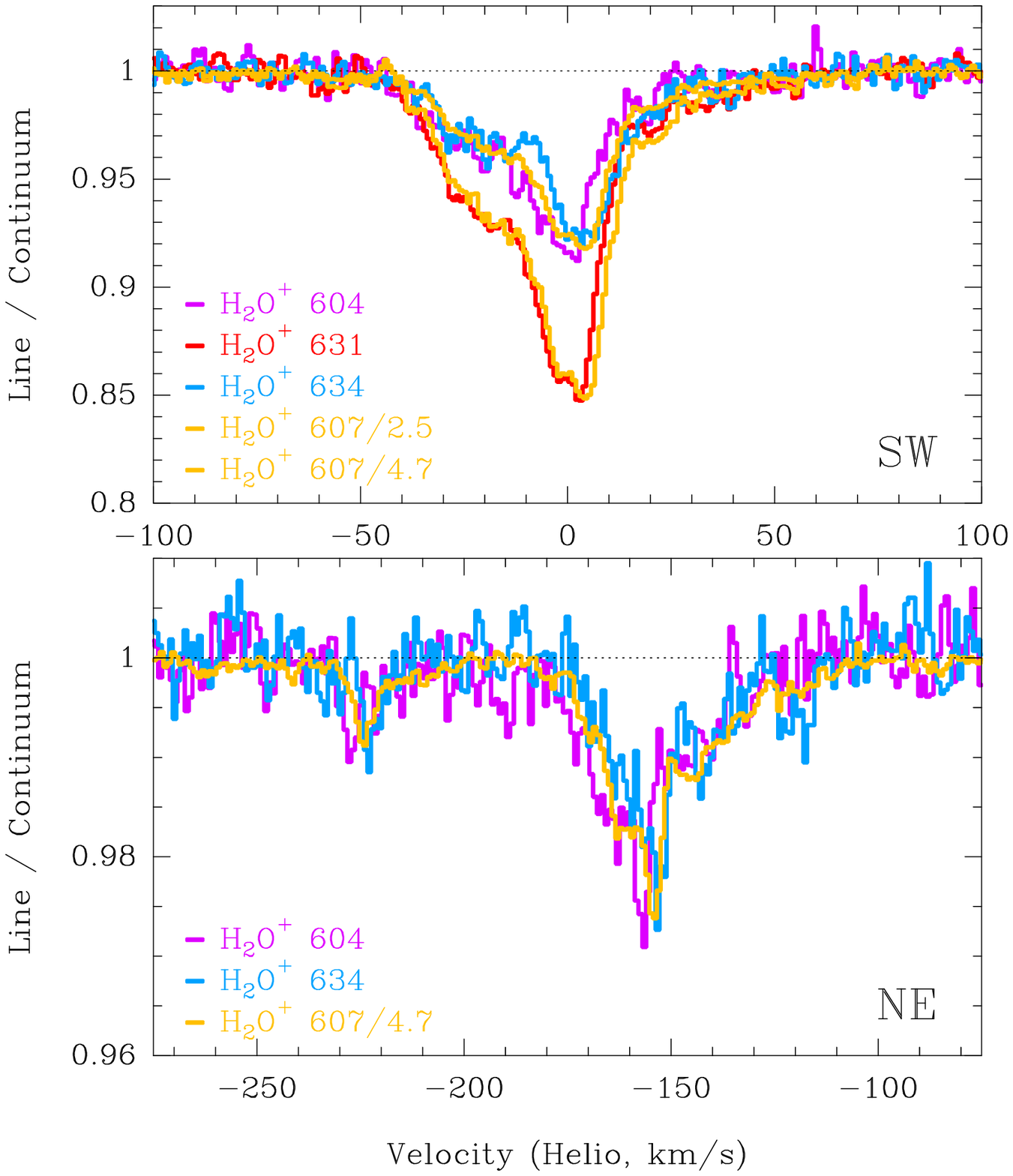}
\includegraphics[width=8.8cm]{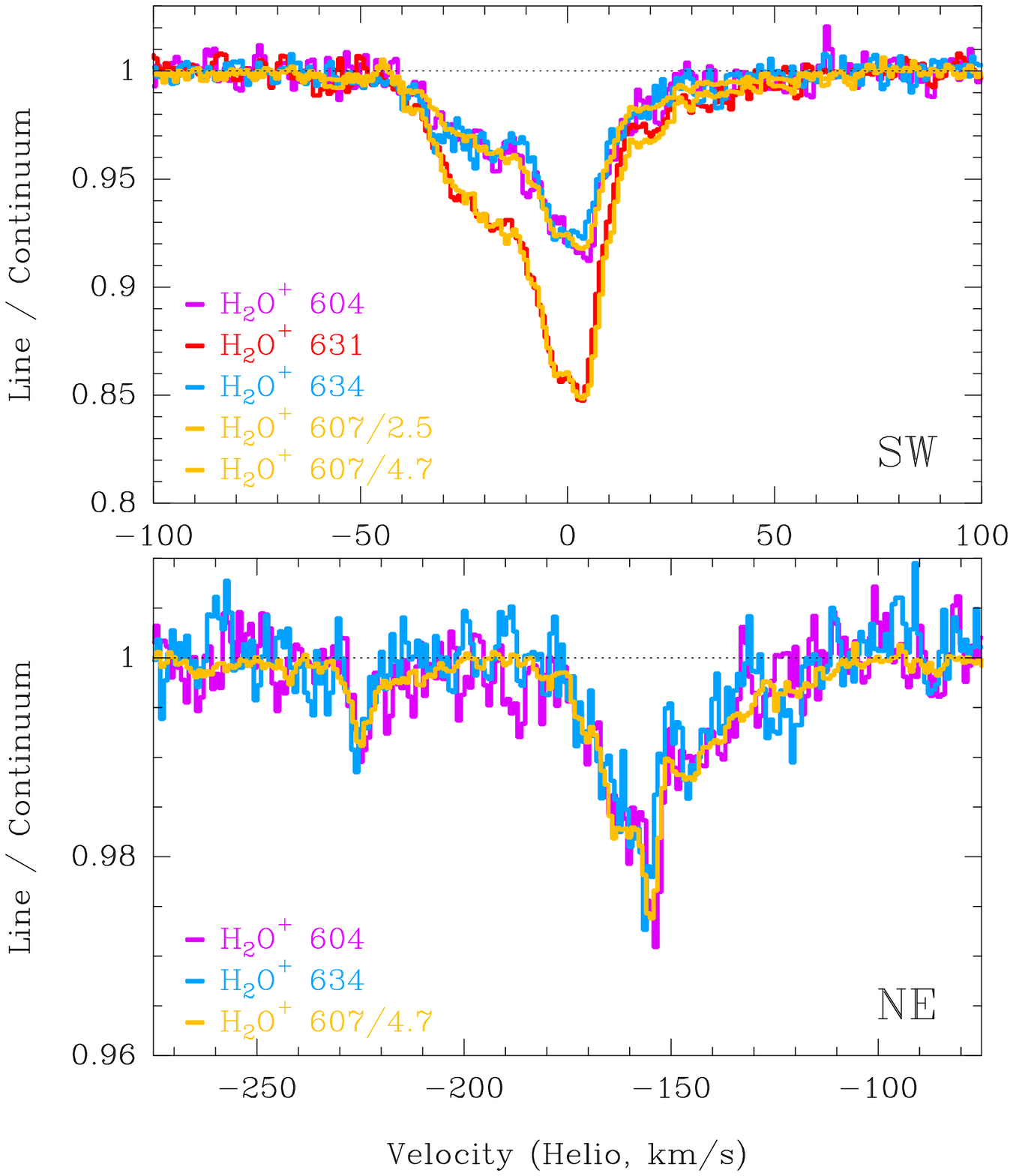}
\caption{Spectra of the four H$_2$O$^+$ lines toward \PKS1830(SW) ({\em top}) and \PKS1830(NE) ({\em bottom}) using current CDMS frequencies ({\em left}) and revised frequencies ({\em right}) as given in Table~\ref{tab:spectro}. The 607~GHz line profile has been scaled down by the ratios of relative strengths.}
\label{fig:spec-H2O+-CDMS-revised}
\end{center} \end{figure*}

The three FS transitions of the {\em p}-H$_2$O$^+$ $1_{10}-1_{01}$ line at 604, 607, and 631~GHz (rest frame) were observed simultaneously in the same tuning for the southwest image. The northeast absorption was out of the band for the 631~GHz transition. The 634~GHz transition was observed approximately one year later (Table~\ref{tab:obs}), as reported by \cite{mul15}, with very minor variations apparent in the line profile compared to the other FS transitions observed in 2014. The optical depths of the four {\em p}-H$_2$O$^+$ FS lines toward the southwest image follow the relative line strengths given in Table~\ref{tab:spectro}, as expected when the sublevels are populated in proportion to their statistical weights (see Fig.\ref{fig:spec-H2O+-CDMS-revised}).

\subsection{Evaluation of H$_2$O$^+$ rest frequencies} \label{sec:evaluateH2O+freq}

The frequencies for the {\em p}-H$_2$O$^+$ lines in the CDMS (version~1) have uncertainties of a few MHz. Using them, we see that our four lines do not match each other precisely in velocity, see Fig.\ref{fig:spec-H2O+-CDMS-revised}(left). We can use our ALMA data to improve the rest frequencies of the observed {\em p}-H$_2$O$^+$ transitions. First, we have cross-correlated the absorption profiles of the {\em p}-H$_2$O$^+$ 607~GHz and {\em o}-H$_2$O 557~GHz lines, which are both optically thin and relatively well correlated toward the NE image. 
The cross-correlation function has a strong peak for an offset (rest frame, taking $z=0.88582$) of 50289.80$\pm$0.17~MHz between the two lines. Using the high-accuracy of 556,935,987.7$\pm$0.3~kHz for the frequency of the {\em o}-H$_2$O line \citep{caz09}, we thus find a frequency of 607225.79$\pm$0.17~MHz for the main FS component of {\em p}-H$_2$O$^+$. In addition, we cross-correlate the four {\em p}-H$_2$O$^+$ FS components to derive their relative frequency offsets, reported in Table\,\ref{tab:cross-corr}. With the new frequencies in Table~\ref{tab:spectro}, the match of the four FS lines is clearly improved with respect to the CDMS frequencies (Fig.\ref{fig:spec-H2O+-CDMS-revised}, right). We note that the relative frequency offsets between {\em p}-H$_2$O$^+$ FS transitions, determined from the cross-correlations, are robust, but that their absolute frequencies are limited by potential chemical segregation and velocity offsets between {\em p}-H$_2$O$^+$ and {\em o}-H$_2$O. This was not taken into account in the frequency uncertainties in Table~\ref{tab:spectro}. More precise and accurate H$_2$O$^+$ frequencies would require laboratory microwave measurements.

\begin{table*}[ht]
\caption{Results for the cross-correlations of the {\em p}-H$_2$O$^+$ lines observed with ALMA.}
\label{tab:cross-corr}
\begin{center} \begin{tabular}{ccc}
\hline \hline
Line pairs & \multicolumn{2}{c}{Frequency shift (MHz)} \\ 
           & CDMS vers.1  & ALMA data \\
\hline

$\Delta$ {\em p}-H$_2$O$^+$(607,604) &  2548.8 $\pm$ 4.3 &  2541.69 $\pm$ 0.58 $^{(a)}$ \\
$\Delta$ {\em p}-H$_2$O$^+$(607,607) &  --               &     0.00 $\pm$ 0.11 $^{(a)}$ \\
$\Delta$ {\em p}-H$_2$O$^+$(631,607) & 24496.8 $\pm$ 5.5 & 24499.29 $\pm$ 0.24 $^{(a)}$ \\
$\Delta$ {\em p}-H$_2$O$^+$(634,607) & 27045.5 $\pm$ 4.2 & 27040.64 $\pm$ 0.61 $^{(a)}$ \\
$\Delta$ {\em p}-H$_2$O$^+$(607)--{\em o}-H$_2$O(557) & 50291.3 $\pm$ 1.9 $^{(c)}$ & 50289.80 $\pm$ 0.17 $^{(b)}$ \\
\hline
\end{tabular} 
\tablefoot{ $a)$ Toward the SW image of \PKS1830; $b)$ Toward the NE image; $c)$ Taking the frequency 556,935,987.7$\pm$0.3~kHz for the ground state line of {\em o}-H$_2$O \citep{caz09}.}
\end{center} \end{table*}

\begin{table*}[ht]
\caption{Spectroscopic parameters for OH$^+$ and {\em p}-H$_2$O$^+$ lines observed in this work.}
\label{tab:spectro}
\begin{center} \begin{tabular}{lcll}
\hline \hline
\multicolumn{1}{c}{Line} & $S_{ul}$ & \multicolumn{1}{c}{$E_{low}$} & \multicolumn{1}{c}{Frequency}  \\
     &         & \multicolumn{1}{c}{(K)} & \multicolumn{1}{c}{(MHz)}          \\
\hline

OH$^+$ $N$=1--0 $J$=0--1 $F$=0.5--0.5 & 0.23 & 0. & 909045.0 (0.8) $^{(a)}$ \\ 
OH$^+$ $N$=1--0 $J$=0--1 $F$=0.5--1.5 & 0.47 & 0. & 909159.4 (0.8) $^{(a)}$ \\ 

\hline
{\em p}-H$_2$O$^+$ $N_{\rm K_a,K_c}$=$1_{1,0}$--$1_{0,1}$ $J$=1.5--0.5  & 0.35 & 0. $^{(b)}$ &  604684.10 (0.76) $^{(c)}$\\
{\em p}-H$_2$O$^+$ $N_{\rm K_a,K_c}$=$1_{1,0}$--$1_{0,1}$ $J$=1.5--1.5 & 1.65 & 0. $^{(b)}$ &  607225.79 (0.17) $^{(c)}$\\
{\em p}-H$_2$O$^+$ $N_{\rm K_a,K_c}$=$1_{1,0}$--$1_{0,1}$ $J$=0.5--0.5 & 0.67 & 0. $^{(b)}$ &  631725.08 (0.42) $^{(c)}$\\
{\em p}-H$_2$O$^+$ $N_{\rm K_a,K_c}$=$1_{1,0}$--$1_{0,1}$ $J$=0.5--1.5 & 0.33 & 0. $^{(b)}$ &  634266.44 (0.78) $^{(c)}$\\

\hline
\end{tabular} 
\tablefoot{$a)$ \cite{mar16};
$b)$ the ground state level of {\em p}-H$_2$O$^+$ is 30~K higher than for the ortho form;
$c)$ from the cross-correlations of ALMA spectra.
}
\end{center} \end{table*}

\subsection{Time variations}

Time variations can be a concern for \PKS1830\ when comparing observations taken at different epochs since the absorption line profiles are known to have variations over a timescale of a few months, related to intrinsic morphological changes in the quasar's jet \citep{mul08,mul14a,sch15}. For example, Very Long Baseline Array {}measurements at 43~GHz by \cite{jin03} revealed that the apparent separation between the NE and SW cores varied by up to 200~$\mu$as, that is, approximately 1.6~pc projected on the plane of the absorber, within 8 months. \cite{nai05} interpreted this apparent motion as possibly due to the episodic ejection of plasmons along a putative helical jet. Accordingly, the two lines of sight through the absorber could vary with time and intercept a different absorbing screen.

\begin{figure}[h] \begin{center}
\includegraphics[width=8.8cm]{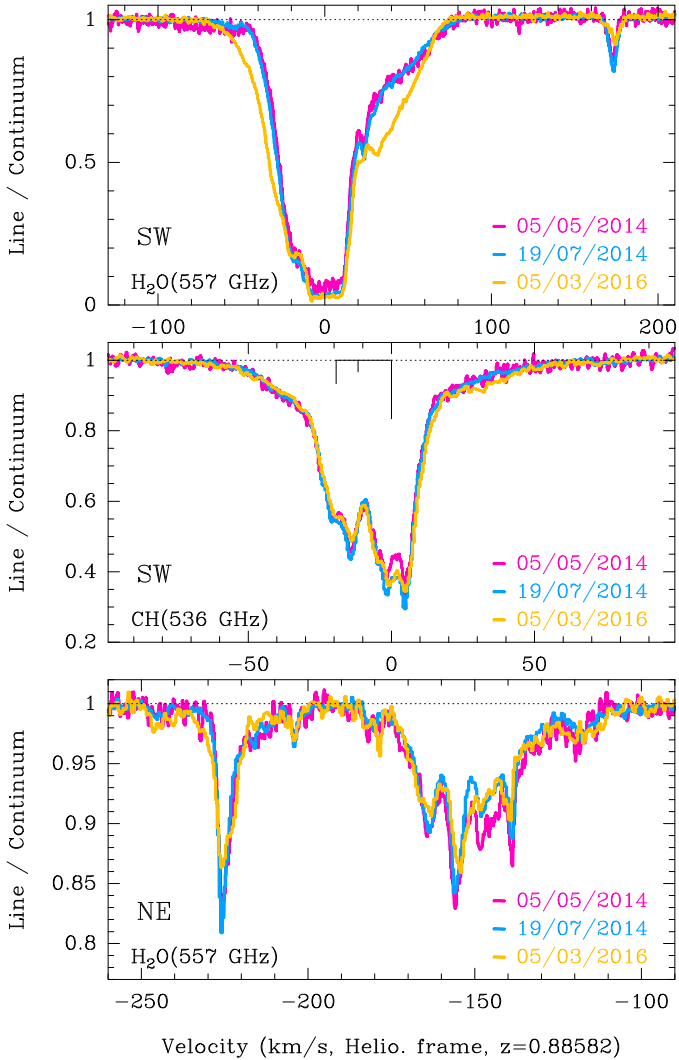}
\caption{Spectra of H$_2$O ({\em top}: toward the southwest image of \PKS1830; {\em bottom}: toward the northeast image) and CH ({\em middle}: toward the southwest image) at several epochs between 2014 and 2016.}
\label{fig:timvar}
\end{center} \end{figure}

Between 2014 and 2016 (the time span of the observations presented here), we have obtained several spectra of the H$_2$O 557~GHz and CH 532/536~GHz lines (Fig.~\ref{fig:timvar}). Toward the southwest image, the CH spectra show only minor changes of less than 6\% of the total integrated opacity. The wings of the saturated line of H$_2$O show the most significant variations in the 2016 spectrum, with an increase of approximately 50\% of the integrated opacities in the velocity ranges $-$20 to $-$60~\kms\ and 20 to 80~\kms, with respect to 2014. Those are small, however, compared to the total integrated opacity over the whole line. Toward the northeast image, the relative variations of the total integrated opacity of the water line (in the velocity interval $-$240 to $-$80~\kms) are less than 15\%. We therefore conclude that time variations do not significantly affect the results presented here.

\subsection{Column densities}

In order to determine the column densities from observations of a single rotational transition of OH$^+$ and H$_2$O$^+$, we must first obtain their excitation temperature. For OH$^+$, the collision rate for destructive reactions with H$_2$ is so close to the elastic collision rate that OH$^+$ formed from exoergic reactions O$^+$+H$_2$ and H$_3^+$+O cannot be fully thermalized during its short lifetime. As a result, OH$^+$ remains kinetically hot compared with other molecules.
At a density of $n({\rm H}_2) \sim 1000$~cm$^{-3}$ (i.e., the order of magnitude estimated toward the southwest image of \PKS1830, \citealt{mul13}), the destruction rate of OH$^+$ is approximately 10$^{-6}$~s$^{-1}$. This is equal to the rate at which OH$^+$ absorbs background radiation in the two 909~GHz lines, 1.08$\times$10$^{-6}$~s$^{-1}$, when the cosmic microwave background (CMB) at $z=0.89$ has $\Tcmb=5.14$~K \citep{mul13}. That is, OH$^+$ is not solely excited by the CMB photons; formation and destruction processes are comparable in importance at a density of the order of 1000~cm$^{-3}$. As long as the destruction rate is not dependent on the FS state, then the excitation temperature will probably closely approximate \Tcmb. The corresponding radiative rates in H$_2$O$^+$ are approximately 20 times higher than the reactive destruction rate at a density of 1000~cm$^{-3}$, therefore, H$_2$O$^+$ is probably more strongly coupled to the CMB. Since we actually expect OH$^+$ and H$_2$O$^+$ to be present in the more diffuse phase, that is, at a density of $< 1000$~cm$^{-3}$, we will assume that their excitation temperature ($T_{ex}$) is locked to the temperature of the cosmic microwave background, $\Tcmb=5.14$~K. If $T_{ex}({\rm OH}^+) > 5.14$~ K, the OH$^+$ column density would rise, and the values determined here would be lower limits. 

In addition, we assume an ortho/para ratio, $OPR$, of 3 for H$_2$O$^+$. \cite{tan13} looked into the ortho to para exchange from the spectroscopic point of view. They found several transitions that could facilitate such an exchange. They concluded, however, that the exchange rates are rather low such that significant alterations of the $OPR$ are unlikely, especially at low temperatures. \cite{sch13} reinvestigated the $OPR$ of H$_2$O$^+$ in the diffuse ISM toward Sgr\,B2(M) with some new observational results and concluded that there is no evidence for an $OPR$ different from the high temperature limit of three.

Toward the southwest image of \PKS1830, we measure integrated opacities $\int \tau dv=122.2\pm0.3$~\kms\ and $13.61\pm0.05$~\kms\ for OH$^+$ (both HFS components together) and {\em p}-H$_2$O$^+$ (only the 607~GHz line), respectively. With the assumptions discussed above, we derive column densities of $1.6\times10^{15}$~\cm-2 and $2.7\times10^{14}$~\cm-2 for OH$^+$ and H$_2$O$^+$, respectively. Toward the northeast image, we measure integrated opacities $\int \tau dv=56.6\pm0.2$~\kms\ and $3.52\pm0.03$~\kms\ for OH$^+$ and {\em p}-H$_2$O$^+$, respectively (over the same HFS/FS components), yielding column densities of a factor 2--4 lower than toward the southwest line of sight (Table~\ref{tab:ncol}).

Among the detected hydrides, OH$^+$, H$_2$O$^+$, as well as H$_2$Cl$^+$ \citep{mul14b} and ArH$^+$ \citep{mul15}, show significantly different line profiles compared to other species, in particular broader line wings and a clear enhancement of their abundance toward the NE image (Table~\ref{tab:ncol}). This is consistent with the expectation that they preferentially trace a diffuse gas component, almost purely atomic in the case of ArH$^+$ \citep{sch14}. The SW/NE column density ratios for these species are between two and four, while for CH and various (non-hydride) species detected in a 7~mm spectral scan with the Australia Telescope Compact Array (ATCA), \cite{mul11} found an average ratio of more than one order of magnitude between the two lines of sight.

The molecular hydrogen fraction derived from the ratio of OH$^+$ and H$_2$O$^+$ is $\fH2 \sim 0.02$--0.04 in both lines of sight (see Sec.\ref{sec:fH2}). This gas component is completely different from that traced by molecules such as HF, H$_2$O, CH, HCN, HCO$^+$, for which the molecular fraction is much higher. More evidence for chemical differentiation along the southwest line of sight comes from the detection of CF$^+$ \citep{mul16}, which has a singular line profile compared to other species observed in the same tuning.

With these ALMA observations of hydrides, we are now starting to identify and disentangle the different gas components in the two absorbing pencil beams in the $z=0.89$ galaxy. The simple comparison of the SW/NE abundance ratios in Table~\ref{tab:ncol} suggests that the sequence of hydrides, ArH$^+$, OH$^+$, H$_2$Cl$^+$, H$_2$O$^+$, CH, and HF, traces gas with an increasing molecular fraction.

\begin{table}[ht]
\caption{Total column densities along the SW and NE lines of sight toward \PKS1830. Species are ordered with increasing SW/NE column density ratios.} \label{tab:ncol}
\begin{center} \begin{tabular}{ccccc}

\hline \hline
Species & \multicolumn{2}{c}{Column densities (\cm-2)} & Ratio & Reference \\
         & SW & NE & SW/NE \\
\hline
H\,I       & $1.3 \times 10^{21}$ & $2.5 \times 10^{21}$ & 0.5 & $(a)$ \\
ArH$^+$     & $2.7 \times 10^{13}$ & $1.3 \times 10^{13}$ & 2.1 & $(b)$ \\
OH$^+$      & $1.6 \times 10^{15}$ & $7.6 \times 10^{14}$ & 2.2 & $(c)$ \\
H$_2$Cl$^+$ & $1.4 \times 10^{13}$ & $3.7 \times 10^{12}$ & 3.8 & $(d)$ \\
H$_2$O$^+$  & $2.7 \times 10^{14}$ & $7.0 \times 10^{13}$ & 3.9 & $(c)$ \\
CH         & $7.7 \times 10^{14}$ & $3.5 \times 10^{13}$ & 22. & $(e)$ \\
HF         & $>3.4 \times 10^{14}$ & $0.18 \times 10^{14}$ & $> 19$ & $(f)$ \\
\hline
H$_2$      &  $\sim 2 \times 10^{22}$ & $\sim 1 \times 10^{21}$ & $\sim 20$ & $(e,g)$ \\
\hline
\end{tabular}
\tablefoot{ $a)$ \cite{koo05}; $b)$ \cite{mul15}; $c)$ This work; $d)$ \cite{mul14b}; $e)$ \cite{mul14a}; $f)$ \cite{kaw16}; $g)$ The column densities of H$_2$ were derived using CH and H$_2$O as proxy, assuming canonical molecular abundances.}
\end{center} \end{table}

\subsection{Fraction of molecular hydrogen} \label{sec:fH2}

With the fraction of molecular hydrogen defined as 

$\fH2=2 \times N({\rm H}_2)/[N({\rm H})+2 \times N({\rm H}_2)]$, 

we need estimates of the total column densities of H\,I and H$_2$ to get average \fH2 values along the absorbing sightlines. The H\,I spectrum toward \PKS1830\ has been obtained by \cite{che99} and \cite{koo05}, although with the radio-cm continuum illumination (NE and SW images embedded in the pseudo-Einstein ring) unresolved. \cite{koo05} did a modeling of the H\,I absorption spectrum, assuming a thin axisymmetric disk in the lensing galaxy and using the quasar radio continuum resolved at higher frequencies. They estimate H\,I opacities of 0.064 and 0.128 toward the southwest and northeast images, respectively, that is, H\,I column densities of $1.3 \times 10^{21} \times \left ( \frac{T_s}{100~K} \right )$~\cm-2\ and $2.5 \times 10^{21} \left ( \frac{T_s}{100~K} \right )$~\cm-2. We further adopt $T_s =100$~K, although it could be marginally higher (e.g., \citealt{dic09,cur16}). Concerning the H$_2$ column densities, we take the values derived by \cite{mul14a} with CH as proxy of H$_2$, assuming the Galactic value [CH]/[H$_2$]=3.5$\times$10$^{-8}$ \citep{she08,ger10}. Accordingly, we obtain $\fH2\sim0.97$ and 0.44 along the SW and NE lines of sight, respectively. These values are somehow uncertain, since the H\,I column densities may have been underestimated. Still, they are consistent with the fact that a large number of molecules is detected toward the almost-purely-molecular $v\sim0$~\kms\ velocity component along the southwest line of sight (e.g., \citealt{mul11}), in contrast to the northeast line of sight which is relatively more diffuse, as indicated by the relative enhancement of OH$^+$, ArH$^+$, H$_2$O$^+$, and H$_2$Cl$^+$.

Since the formation and destruction of OH$^+$ and H$_2$O$^+$ are controlled by simple abstraction reactions with H$_2$ and dissociative recombinations with electrons, their abundance ratio can be used as a local measurement of the fraction of molecular gas, as shown by \cite{hol12}. The analytical expression of \fH2\ (see Eq.12 from \citealt{ind15}) scales linearly with the fractional abundance of electrons, \xe, and also depends on the gas kinetic temperature, \Tkin, via the rate coefficients for dissociative recombination of OH$^+$ and H$_2$O$^+$ with electrons (taken from the UMIST Database for Astrochemistry 5th release; \citealt{mcelr13}). \xe\ cannot be constrained from our data and so, we adopt the same value as \cite{ind15}, $\xe=1.5 \times 10^{-4}$, which comes from C$^+$ observations in the Milky Way. A different metallicity in the $z=0.89$ absorber (which is yet unknown) could potentially affect \xe; however, the chemical abundances of the molecular species observed toward \PKS1830\ are comparable to those in Galactic diffuse and/or translucent clouds (\citealt{mul11}). For the kinetic temperature, we assume a canonical value of 100~K, close to that determined by \cite{mul13} toward the SW image, and identical to that taken by \cite{ind15} for all their Galactic clouds. With these assumptions for \xe\ and \Tkin\ and the OH$^+$/H$_2$O$^+$ ratios from Table~\ref{tab:ncol}, we obtain $\fH2=0.04$ in the southwest line of sight and $\fH2=0.02$ in the northeast one. Varying the kinetic temperature in the range 30--300~K would only change \fH2\ between 0.02--0.07 and 0.01--0.04, for the SW and NE lines of sight, respectively.

These values are much lower than the average \fH2 derived above, indicating that OH$^+$ and H$_2$O$^+$ are found in a gas component with a very low H$_2$ fraction. OH$^+$ and H$_2$O$^+$ provide the H$_2$ fraction in the regions where they reach their maximum abundance, which can be the regions where H\,I is the main hydrogen species. The similarity (only a factor of two) between the molecular fractions \fH2\ derived from OH$^+$/H$_2$O$^+$ along the NE and SW lines of sight, despite the known difference in their cloud type populations and in galactocentric distance, is remarkable, as well as the similarity with \fH2\ values obtained by \cite{ind15} in the Milky Way. It may be related to the fact that OH$^+$ and H$_2$O$^+$ trace a low-\fH2, diffuse gas regime which is always present along these lines of sight. The difference in mean molecular fraction appears more clearly in other tracers, more sensitive to the bulk of molecular gas, such as CH or H$_2$O. Indeed, the difference in line profile between these species toward \PKS1830\ shows that they are not fully mixed (see also the results of the Principal Component Analysis of the Galactic line of sight toward W31C presented by \citealt{ger16}, their Fig.~6). The profile of the H$_2$O$^+$ line has broader wings than that of the H$_2^{18}$O line, which is expected to trace the densest clouds, but is narrower than the ArH$^+$ line, which traces the almost purely atomic component. This forms a coherent scenario where multi-phase gas components have different scale heights in the disk of the lensing galaxy that is seen approximately face-on.

\subsection{Cosmic-ray ionization rate of atomic hydrogen}

The abundance ratio of OH$^+$ and H$_2$O$^+$ can also be used to infer the cosmic-ray ionization rate of hydrogen, \zzH\, \citep{hol12, ind15}. The analytical expression of \zzH\ (Eq.15 from \citealt{ind15}) requires \fH2\ and \xe, as well as the H\,I column density (Table~\ref{tab:ncol}) and the volume density of hydrogen, \nH. This latter quantity cannot be constrained from our data, and so we adopt the same value as \cite{ind15}, $n_{\rm H}=35$~cm$^{-3}$ (resulting from a pressure equilibrium statement) for both NE and SW lines of sight, as well as the same efficiency factor $\epsilon=0.07$ as \cite{ind15}, which takes into account the fact that not all hydrogen ionization results in the formation of OH$^+$.

\begin{figure}[h] \begin{center}
\includegraphics[width=8.8cm]{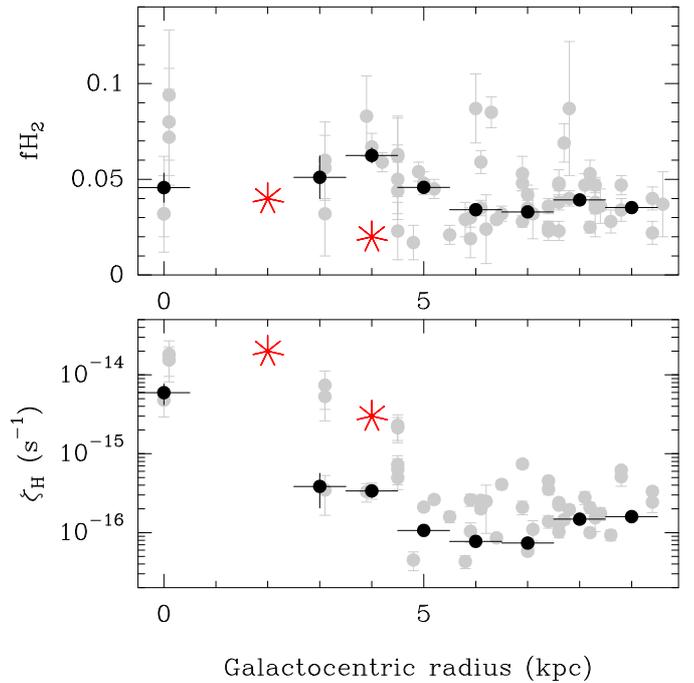}
\caption{Molecular hydrogen fraction (\fH2, {\em top}) and cosmic-ray ionization rate (\zzH, {\em bottom}) vs galactocentric radius in the $z=0.89$ absorber toward \PKS1830\ ({\em the two red stars}) and in the Milky Way ({\em light gray dots}: foreground clouds only, taken from \citealt{ind15}; {\em black dots}: same Galactic entries, but weighted-averaged over 1~kpc bins).}
\label{fig:Indriolo15}
\end{center} \end{figure}

With these values, we obtain $\zzH \sim 2 \times 10^{-14}$~s$^{-1}$ and $\sim 3 \times 10^{-15}$~s$^{-1}$ for the southwest and northeast lines of sight, respectively, that is, a decrease by a factor of six between 2 and 4~kpc (assuming that \xe\ and \nH\ are the same). This gradient of \zzH\ is consistent with that observed in the Milky Way \citep{ind15}. If we roughly interpolate between the few Galactic measurements by \cite{ind15} at galactocentric radii 0--5~kpc, we find that our values of \zzH\ in the $z=0.89$ absorber are slightly higher than in the Milky Way (Fig.~\ref{fig:Indriolo15}), although this could be due to uncertainties (assumptions of \xe, \nH, covering factor, co-spatiality of species) and/or the small number of measurements in Galactic clouds in the inner 5~kpc. We speculate that a higher \zzH\ in the $z=0.89$ absorber could be related to a higher average star formation activity than in the Milky Way. Indeed, the differences in elemental isotopic ratios between the $z=0.89$ absorber and the Milky Way (e.g., $^{32}$S/$^{34}$S, \citealt{mul06}; $^{36}$Ar/$^{38}$Ar, \citealt{mul15}) already implied that the former experienced phases of high mass star formation.

One possible caveat as pointed out by \cite{hol12}: when the ionization rate is extreme in Galactic terms, $\zzH \geq 10^{-15}$~s$^{-1}$, the electron abundance grows beyond the carbon abundance limit owing to ionization of hydrogen. As a result, OH$^+$ and H$_2$O$^+$ molecules can be destroyed mainly by electrons rather than H$_2$, favoring OH$^+$, with less dependence on the molecular fraction. With such high ionization rates, it would no longer be appropriate to assume an electron fraction as low as $1.5 \times 10^{-4}$ and the estimates of \fH2\ and \zzH\ from the abundance ratio of OH$^+$ and H$_2$O$^+$ may, in fact, be underestimates, as shown by \cite{lep16} using additional H$_3^+$ data and photo-dissociation region models in the central molecular zone in the Milky Way.

\section{Summary and conclusions} \label{sec:conclusions}

We present ALMA observations of OH$^+$ and H$_2$O$^+$ along two independent sight lines in the $z=0.89$ absorber toward the quasar \PKS1830. We find:
\begin{itemize}
\item OH$^+$ and H$_2$O$^+$ are both detected along the two lines of sight and, by comparison with other hydrides, show evidence for a multi-phase gas composition in the absorbing gas.
\item The column density ratios between the SW and NE lines of sight increase in a sequence ArH$^+$, OH$^+$, H$_2$Cl$^+$, H$_2$O$^+$, CH, and HF, consistent with our understanding that these hydrides trace gas with an increasing molecular fraction.
\item The abundance ratios between OH$^+$ and H$_2$O$^+$ are used to infer the fraction of molecular hydrogen and cosmic-ray ionization rate in the disk of the absorber. We find values ($\fH2=0.04$ and 0.02; $\zzH \sim 2 \times 10^{-14}$~s$^{-1}$ and $\sim 3 \times 10^{-15}$~s$^{-1}$ for the southwest and northeast lines of sight, respectively) which are comparable to those in the Milky Way at similar galactocentric radii ($\sim 2$~kpc and $\sim 4$~kpc, respectively). The slightly higher ionization rates found in the $z=0.89$ absorber could be related to a higher average star formation activity than in the Milky Way.
\item We revise the rest frequencies of H$_2$O$^+$ lines, although precise laboratory microwave measurements are required to further increase the accuracy.
\end{itemize}

The number of sources for which observations of OH$^+$ and H$_2$O$^+$ have been reported and where \fH2 and \zzH\ have been estimated is still limited. The ALMA observations of \PKS1830\ presented in this paper allow us to investigate \fH2 and \zzH\ in the disk of a distant $z=0.89$ galaxy, and compare them to those in the Milky Way. With the discovery of new high-redshift molecular absorbers, such studies could be drastically extended with ALMA.

\begin{acknowledgement}
We thank the anonymous referee for her/his useful comments which helped us to improve the clarity of the manuscript.
This paper makes use of the following ALMA data:
ADS/JAO.ALMA\#2012.1.00056.S, \#2013.1.00020.S, and \#2015.1.00075.S.
ALMA is a partnership of ESO (representing its member states), NSF (USA) and NINS (Japan), together with NRC (Canada) and NSC and ASIAA (Taiwan) and KASI (Republic of Korea), in cooperation with the Republic of Chile. The Joint ALMA Observatory is operated by ESO, AUI/NRAO and NAOJ.
\end{acknowledgement}

\appendix

\section{Laboratory spectroscopic details of H$_2$O$^+$} \label{sec:LabSpectroH2O+}


\begin{table*}
\begin{center}
\caption{Rest frequencies$^a$ (MHz) of \textit{para}- and \textit{ortho}-H$_2$O$^+$ 
         ground state transitions from different sources.}
\label{H2O+_freqs}
\renewcommand{\arraystretch}{1.10}
\begin{tabular}[t]{lr@{}lr@{}lr@{}lr@{}l}
\hline \hline
Quanta$^b$ & \multicolumn{2}{c}{ALMA$^c$} & \multicolumn{2}{c}{CDMS new$^c$} & 
 \multicolumn{2}{c}{CDMS old$^d$} & \multicolumn{2}{c}{\cite{mur98}$^e$} \\ 
\hline
 \multicolumn{9}{l}{ } \\
 \multicolumn{9}{l}{\textit{para}-H$_2$O$^+$, $1_{10} - 1_{01}$, $J =$} \\
\hline
$1.5 - 0.5$ & 604684&.10~(76) &  604682&.66~(38) &  604678&.6~(25) &  604758&.7 \\
$1.5 - 1.5$ & 607225&.79~(17) &  607225&.80~(17) &  607227&.3~(19) &  607223&.9 \\ 
$0.5 - 0.5$ & 631725&.08~(42) &  631724&.98~(39) &  631724&.1~(37) &  631773&.0 \\
$0.5 - 1.5$ & 634266&.44~(78) &  634268&.11~(39) &  634272&.9~(24) &  634238&.2 \\
 \multicolumn{9}{l}{ } \\
 \multicolumn{9}{l}{\textit{ortho}-H$_2$O$^+$, $1_{11} - 0_{00}$, $J = 1.5 - 0.5$, $F =$} \\
\hline
$1.5 - 0.5$ &       &         & 1115155&.98~(76) & 1115150&.75~(85) & 1115150&.3 \\
$0.5 - 0.5$ &       &         & 1115191&.40~(72) & 1115186&.18~(81) & 1115186&.0 \\
$2.5 - 1.5$ &       &         & 1115209&.33~(73) & 1115204&.15~(82) & 1115204&.0 \\
$1.5 - 1.5$ &       &         & 1115268&.08~(75) & 1115262&.90~(82) & 1115263&.0 \\
$0.5 - 1.5$ &       &         & 1115303&.50~(82) & 1115298&.33~(87) & 1115298&.7 \\
 \multicolumn{9}{l}{ } \\
 \multicolumn{9}{l}{\textit{ortho}-H$_2$O$^+$, $1_{11} - 0_{00}$, $J = 0.5 - 0.5$, $F =$} \\
\hline
$0.5 - 0.5$ &       &     & 1139536&.62~(96)$^f$ & 1139541&.5~(10)  & 1139541&.1 \\
$1.5 - 0.5$ &       &     & 1139555&.79~(84)$^f$ & 1139560&.6~(9)   & 1139560&.6 \\
$0.5 - 1.5$ &       &     & 1139648&.73~(85)$^f$ & 1139653&.7~(9)   & 1139653&.5 \\
$1.5 - 1.5$ &       &     & 1139667&.89~(91)$^f$ & 1139672&.7~(10)  & 1139673&.3 \\
 \multicolumn{9}{l}{ } \\
 \multicolumn{9}{l}{\textit{para}-H$_2$O$^+$, $2_{12} - 1_{01}$, $J =$} \\
\hline
$2.5 - 1.5$ &       &         & 1626333&.5~(13)  & 1626324&.6~(23)  & 1626315&.0 \\
$1.5 - 0.5$ &       &         & 1638944&.9~(12)  & 1638934&.8~(20)  & 1638972&.0 \\
$1.5 - 1.5$ &       &         & 1641488&.1~(12)  & 1641483&.6~(24)  & 1641437&.2 \\
\hline
\end{tabular} 
\end{center}
\tablefoot{
$^a$ Numbers in parentheses after the values refer to the uncertainties 
     in units of the least significant figures.
$^b$ Final quantum numbers $J' - J''$ (\textit{para}) or $F' - F''$ (\textit{ortho}). 
$^c$ This work.
$^d$ Version 1, available, e.g., at 
http://www.astro.uni-koeln.de/site/vorhersagen/catalog/archive/H2O+/vers.1/.
$^e$ Estimated uncertainties are about 1.8~MHz throughout \cite{mur98}.
$^f$ Tentative, see text.
}
\end{table*}


The initial CDMS entry (version 1) was derived using extrapolated field-free transition frequencies from the laser magnetic resonance (LMR) study of \cite{mur98}. These authors also used data from the earlier LMR study of \cite{str86}, but only reported the calculated frequencies for the $1_{11}$--$0_{00}$ transitions. The data set also included infrared (IR) data from \cite{din88} ($\nu _3$), \cite{bro89} ($\nu _2$), \cite{hue92} ($\nu _1$ and $\nu _3$), and \cite{zhe08} ($\nu _2$). Furthermore, it included ground state combination differences (GSCDs) from the $A - X$ electronic spectrum in the near-infrared \citep{lew76,hue97}. Spectroscopic parameters were determined using Pickett's SPFIT program \citep{pic91} which can only deal with field-free data, as is the case for most spectroscopy programs. Starting parameters were taken from \cite{mur98} for the ground vibrational state and from \cite{hue92} and \cite{zhe08} for the excited vibrational states. Using considerably more IR data than \cite{mur98}, they only used GSCDs from \cite{din88}, and, in addition, extensive GSCDs derived from \cite{hue97}, but we could only use a small part of the \cite{str86} data, as the set of spectroscopic parameters differed somewhat from the initial set of parameters. Some changes were caused by using common parameters for all or most states for some parameters which were only poorly determined (mostly parameters of a higher order).

Although our ALMA data for H$_2$O$^+$ can be modeled much better with the old CDMS data than with the frequencies derived from \cite{mur98}, the model was not satisfactory. Furthermore, \cite{neu10} suggested from observations of the OH$^+$ 1$-$0 transition and of \textit{ortho}-H$_2$O$^+$ near 1115~GHz with Herschel that the rest frequencies of the latter cation should be shifted up in frequency by 5~MHz. We decided to use these data to refine the spectroscopic parameters of H$_2$O$^+$. The new or modified frequencies were accomodated quite well. Having shifted the $J$=1.5--0.5 group of lines of the $1_{11}$--$0_{00}$ transition near 1115~GHz, the question remained of how to treat the $J$=0.5--0.5 group of lines near 1140~GHz. Shifting them also up by 5~MHz deteriorated the fit marginally and reproduced the $\Delta J$=$\pm$1 FS lines of the $1_{10}$--$1_{01}$ transition less well by approximately 0.6~MHz. Shifting the 1140~GHz group of lines down by 5~MHz improved the fit marginally and improved reproduction of the $\Delta J$=$\pm$1 FS lines of the $1_{10}$--$1_{01}$ transition by about 0.6~MHz. The 1140~GHz group of lines does not shift appreciably from the positions reported by \cite{mur98} if they were weighted out of the fit. Therefore, we propose tentatively that frequencies of the 1140~GHz group of lines may be somewhat lower (around 5~MHz, possibly more) than the frequencies reported by \cite{mur98}. Additional data, preferably from laboratory measurements with microwave accuracy, should be able to resolve the frequency issues of these lowest energy transitions. Until then, we recommend our ALMA rest frequencies for the $1_{10}$--$1_{01}$ transition and the newly calculated ones for all other transitions.

The number of independent spectroscopic parameters did not change with respect to those used for version 1 of the CDMS entry of H$_2$O$^+$. However, two ground state parameters ($\phi _K$ and $\phi _{NK}$) were replaced by two others (${\it \Phi}_{KN}$ and $L_{KKN}$). We report the resulting spectroscopic parameters in Table~\ref{H2O+_spec-param} and will provide the line, parameter, and fit files in the CDMS catalog archive\footnote{http://www.astro.uni-koeln.de/site/vorhersagen/catalog/archive/H2O+/}. The intensities were calculated using the quantum chemically calculated value of the dipole moment from \cite{wei89} as previously.


\begin{table*}
\begin{center}
\caption{Spectroscopic parameters (cm$^{-1}$, MHz)$^a$ of H$_2$O$^+$ in the ground and excited vibrational states.}
\label{H2O+_spec-param}
\begin{tabular}[t]{lr@{}lr@{}lr@{}lr@{}l}
\hline \hline
Parameter                     & \multicolumn{2}{c}{$\varv = 0$}             & \multicolumn{2}{c}{$\varv _2 = 1$} 
                              & \multicolumn{2}{c}{$\varv _1 = 1$}          & \multicolumn{2}{c}{$\varv _3 = 1$} \\
\hline
$E_{\rm vib}$                 &        0&.0           &     1408&.4131~(20) &     3212&.8567~(26) &     3259&.0341~(11) \\
$A$                           &   870580&.8~(5)       &  1001285&.~(35)     &   851254&.~(61)     &   835041&.~(13)     \\
$B$                           &   372365&.4~(9)       &   374077&.4~(94)    &   365511&.7~(165)   &   367803&.7~(79)    \\
$C$                           &   253880&.4~(5)       &   249275&.7~(59)    &   248680&.5~(84)    &   249733&.7~(46)    \\
$F_{ab}$(13)                  &         &             &         &           &    13273&.~(12)     &         &           \\
${\it \Delta}_K$              &     1375&.3~(6)       &     2902&.6~(60)    &     1348&.5~(145)   &     1269&.2~(14)    \\
${\it \Delta}_{NK}$           &   $-$155&.30~(22)     &   $-$246&.4~(19)    &   $-$154&.6~(15)    &   $-$158&.6~(7)     \\
${\it \Delta}_N$              &       29&.66~(3)      &       31&.00~(19)   &       29&.80~(21)   &       30&.23~(12)   \\
$\delta _K$                   &       51&.81~(11)     &      134&.58~(96)   &       52&.90~(179)  &       49&.77~(101)  \\
$\delta _N$                   &       11&.346~(15)    &       12&.599~(51)  &       11&.471~(112) &       11&.620~(59)  \\
${\it \Phi}_K$                &        7&.95~(8)      &       32&.18~(27)   &        9&.39~(105)  &        7&.40~(8)    \\
${\it \Phi}_{KN} \times 10^3$ &   $-$327&.~(29)$^b$   &         &           &         &           &         &           \\
${\it \Phi}_{NK} \times 10^3$ &   $-$229&.~(7)$^c$    &  $-$1059&.~(32)     &         &           &         &           \\
${\it \Phi}_N \times 10^3$    &       14&.3~(7)$^b$   &         &           &         &           &         &           \\
$\phi _N \times 10^3$         &        6&.74~(32)$^b$ &         &           &         &           &         &           \\
$L_K \times 10^3$             &    $-$79&.0~(35)$^b$  &         &           &         &           &         &           \\
$L_{KKN} \times 10^3$         &       11&.3~(13)$^b$  &         &           &         &           &         &           \\
$P_K \times 10^3$             &        0&.22$^d$      &         &           &         &           &         &           \\
$\epsilon_{aa}$               & $-$32610&.6~(11)      & $-$48469&.~(76)     & $-$32360&.~(164)    & $-$30515&.~(37)     \\
$\epsilon_{bb}$               &  $-$3435&.4~(8)       &  $-$3648&.~(19)     &  $-$3385&.~(32)     &  $-$3425&.~(20)     \\
$\epsilon_{cc}$               &       41&.6~(8)$^b$   &         &           &         &           &         &           \\
${\it \Delta}^S_K$            &      228&.0~(11)      &      505&.1~(77)    &      269&.7~(198)   &      209&.3~(22)    \\
${\it \Delta}^S_{NK}$         &    $-$19&.8~(5)$^b$   &         &           &         &           &         &           \\
${\it \Delta}^S_N$            &        0&.95~(4)$^b$  &         &           &         &           &         &           \\
$\delta ^S_N$                 &        0&.56~(4)$^b$  &         &           &         &           &         &           \\
${\it \Phi}^S_K$              &     $-$2&.21~(4)$^e$  &         &           &         &           &         &           \\
$a_F$(H)                      &    $-$74&.73~(29)     &         &           &         &           &         &           \\
$T_{aa}$(H)                   &       39&.5~(8)       &         &           &         &           &         &           \\
$T_{bb}$(H)                   &    $-$18&.3~(6)       &         &           &         &           &         &           \\
$|T_{ab}(H)|$                 &       72&.$^f$        &         &           &         &           &         &           \\
\hline
\end{tabular}
\end{center}
\tablefoot{
$^a$ All units are in MHz except $E_{\rm vib}$ in cm$^{-1}$. Numbers in parentheses are one standard deviation in units 
     of the least significant figures. Parameters without quoted uncertainties were estimated and kept fixed in the analyses. 
$^b$ Common parameter for all states. 
$^c$ Common parameter for all states except for $\varv _2 = 1$. 
$^d$ Estimated common parameter for all states. 
$^e$ Common parameter for all states except for $\varv _2 = 1$; $\varv _2 = 1$ value constrained to be two times the ground 
     state value. 
$^f$ From \cite{sta93}, quantum chemically calculated. 
}
\end{table*}

\end{document}